\documentclass[superscriptaddress,aps,prd,showpacs, twocolumn]{revtex4}

\def\be{\begin{equation}}
\def\ee{\end{equation}}
\def\bea{\begin{eqnarray}}
\def\eea{\end{eqnarray}}

\begin{document}
\title{Wormhole geometries supported by quark matter at ultra-high densities}

\author{Tiberiu Harko}
\email{t.harko@ucl.ac.uk}
\affiliation{Department of Mathematics, University College London, Gower Street, London
WC1E 6BT, United Kingdom}
\author{Francisco S. N. Lobo}
\email{flobo@cii.fc.ul.pt}
\affiliation{Centro de Astronomia e Astrof\'{\i}sica da Universidade de Lisboa, Campo
Grande, Edific\'{\i}o C8, 1749-016 Lisboa, Portugal}
\author{M. K. Mak}
\email{mankwongmakk@gmail.com}
\affiliation{Department of Computing and Information Management, Hong Kong Institute of
Vocational Education, Chai Wan, Hong Kong, P. R. China}
\date{\today}

\begin{abstract}

A fundamental ingredient in wormhole physics is the presence of
exotic matter, which involves the violation of the null energy
condition. In this context, we investigate the possibility that
wormholes could be supported by quark matter at extreme densities.
Theoretical and experimental investigations of the structure of
baryons show that strange quark matter, consisting of the $u$,
$d$ and $s$ quarks, is the most energetically favorable state of
baryonic matter. Moreover, at ultra-high densities, quark matter may exist in a variety of
superconducting states, namely, the Color-Flavor-Locked (CFL) phase. Motivated by these theoretical models, we explore the conditions under which wormhole geometries
may be supported by the equations of state considered in the
theoretical investigations of quark-gluon interactions. For the description of the normal quark matter we adopt the Massachusetts Institute of Technology (MIT) bag model equation of state, while the color superconducting quark phases are described by a first order approximation of the free energy. By assuming specific forms for the bag and gap functions, several wormhole models are obtained for both normal and superconducting quark matter. The effects of the presence of an electrical charge are also taken into account.

\end{abstract}


\date{\today}

\maketitle

\section{Introduction}

A fundamental property in wormhole physics, in the context of
classical general relativity, is that these exotic geometries are
supported by ``exotic matter''  \cite{Morris}, which involves a
stress-energy tensor $T_{\mu\nu}$ that violates the null energy condition
(NEC), i.e., has $T_{\mu\nu}k^{\mu}k^{\nu}<0$ at the wormhole throat and its neighbourhood, where $k^{\mu}$ is {\it any} null vector \cite{Morris,Visser}. A wide variety of solutions
have been obtained since the seminal Morris-Thorne paper
\cite{Morris}, ranging from dynamic wormhole geometries
\cite{dynWH}, rotating solutions \cite{Teo:1998dp}, thin-shell
wormholes constructed using the cut-and-paste technique
\cite{thinshell}, observational signatures using thin accretion
disks \cite{Harko:2008vy}, solutions in conformal symmetry, which
presents a more systematic approach in searching for exact
wormhole solutions \cite{Boehmer:2007rm}, wormhole geometries in
the semi-classical regime \cite{semiclassWH}, and more recently in
modified theories of gravity \cite{modgrav,modgrav2}.

In the modified gravity context, it was shown that the normal matter
threading the wormhole can be constrained to satisfy the null energy
condition, and it is the higher order curvature terms, interpreted
as a gravitational fluid, that sustain these non-standard wormhole
geometries, fundamentally different from their counterparts in
general relativity. It has also been argued that wormhole
solutions can be supported by several dark energy models
responsible for the late-time cosmic acceleration
\cite{phantomWH}, by imposing specific equations of state. In this
work, we explore the possibility that wormholes could be supported by
quark matter at extreme densities.

This approach is motivated by theoretical and experimental
investigations of baryonic structure showing that strange quark
matter, consisting of the $u$ (up), $d$ (down) and $s$ (strange) quarks is the most
energetically favorable state of baryon matter. The idea of the existence of stars made of quarks was initially introduced in \cite{It} and \cite{Bod}.
Two ways of formation of stellar  strange matter have been proposed in \cite{1} and \cite{2,2c}: the
quark-hadron phase transition in the early universe, and the conversion
of neutron stars into strange ones at ultrahigh densities. In the theories of strong interactions the quark
bag models suppose that the
breaking of physical vacuum takes place inside hadrons. As a
result the vacuum energy densities inside and outside a hadron
become essentially different and the vacuum pressure $B$ on a bag
wall equilibrates the pressure of quarks thus stabilizing the
system \cite{2,2c}.

The structure of a realistic strange star is very complicated but
its basic properties can be described as follows \cite{2,2c}.
Beta-equilibrated strange quark-star matter consists of an
approximately equal mixture of $u$, $d$ and $s$ quarks, with a
slight deficit of the latter. The Fermi gas of $3A$ quarks
constitutes a single color-singlet baryon with baryon number $A$.
This structure of the quarks leads to a net positive charge inside
the star. Since stars in their lowest energy state are supposed to
be charge neutral, electrons must balance the net positive quark
charge in strange matter stars \cite{2,2c}.

However, the electrons, being bound to the quark matter by the
electromagnetic interaction only (and not by the strong force),
are able to displace freely across the quark surface. But they
cannot move to infinity because of the electrostatic interaction
with quarks.
The electron distribution extends up to $\sim 10^{3}$ fm above the
quark surface. The Coulomb barrier at the quark surface of a hot
strange star could represent a powerful source of
electron-positron ($e^{+}e^{-}$) pairs
\cite{Us98}, which are created in the extremely strong electric field of
the barrier. At surface temperatures of around $10^{11}$ K, the
luminosity of the quark star surface  may be of the order $\sim
10^{51}$ ergs$^{-1}$ \cite{elpos}. Moreover,  due to both photon emission and $e^{+}e^{-}$ pair
production, for about $8.6\times 10^4$ s
for normal quark matter and for up to around $3\times 10^9$ s for
superconducting quark matter,  the thermal luminosity from the quark star
surface may be orders of magnitude higher than the Eddington
limit \cite{PaUs02}.

The existence of a large variety of color superconducting states
of quark matter at ultra-high densities has also been suggested and
intensively investigated \cite{Al1,Al2,Al3,Al4}. At very high
densities, matter is expected to form a degenerate Fermi gas of
quarks in which the quark Cooper pairs with very high binding
energy condense near the Fermi surface. This phase of the quark matter is called a  color superconductor.
Such a state is significantly more bound than ordinary quark
matter. This implies that at extremely high density
the ground state of quark matter is the superconducting
Color-Flavor-Locked (CFL) phase, and that this phase of matter
rather than nuclear matter may be the ground state of hadronic
matter \cite{Al4}. The existence of the CFL phase can enhance the
possibility of the existence of a pure stable quark star \cite{Al4}.

In this context, the possibility that stellar mass black holes,
with masses in the range of $3.8M_{\odot}$ and $6M_{\odot}$,
respectively, could be in fact quark stars in the
CFL phase was considered in \cite{Zoltan}.
Depending on the value of the gap parameter, rapidly rotating CFL
quark stars can achieve much higher masses than standard neutron
stars, thus making them possible stellar mass black hole
candidates. Moreover, quark stars have a very low luminosity and a
completely absorbing surface -- the infalling matter on the
surface of the quark star is converted into quark matter.

It is the purpose of the present paper to investigate the
possibility that  wormhole geometries can be realized by using
quark matter, in both normal and superconducting phases. To describe quark matter we adopt the Massachusetts Institute of Technology (MIT) bag model equation of state, while for the investigation of the superconducting quark matter we consider the equation of state obtained in a first order expansion of the free energy of the system. Generally the equations of state depend on several parameters, of which the most important are the bag and the gap constant. The bag constant forces the quarks to remain confined inside the baryons, while the gap constant describes the superconducting properties of the quark matter. However, in high density systems, which can be achieved, for example, in the interior of neutron stars, both the bag and the gap constants, as well as the quark masses, become effective, density dependent functions. It is exactly this property of strongly interacting systems in dense media we will exploit in order to obtain wormhole solutions of the static, spherically symmetric gravitational field equations in the presence of quark matter. By appropriately choosing the forms of the bag and gap functions several wormhole type solutions of the gravitational field equations are obtained, with the matter source represented by normal and superconducting quark matter, respectively.

The present paper is organized as follows. In Section~\ref{eos},
the quark matter equations of state are presented. In
Section~\ref{secII}, we explore the conditions under which wormhole
geometries may be supported by the equations of state considered
in the theoretical investigations of quark-gluon interactions. We discuss and conclude our results in Section \ref{concl}

\section{Quark matter equations of state}\label{eos}

The state of matter at extreme densities represents one of the
most important subjects of study in present day physics. The
problem is complicated, not only from the theoretical point of
view, but also by the fact that laboratory experiments cannot
provide the necessary data for a full understanding of the
question. In order to test our understanding of the relevant
physics we need to turn to astrophysics, and the dynamics of
compact general relativistic objects. In fact, ``neutron stars''
represent unique laboratories of such extreme physics. With core
densities reaching about one order of magnitude beyond nuclear
saturation, they are likely to contain exotic states of matter
like hyperon phases with net strangeness and/or deconfined quarks \cite{Weber}.

The theory of the equation of state of quark matter is directly
based on the fundamental Quantum Chromodynamics (QCD) Lagrangian, given by \cite{3}
\bea
L_{QCD}&=&\frac{1}{4}\sum_{a}F_{\mu \nu }^{a}F^{a\mu \nu
}+\nonumber\\
&&\sum_{f=1}^{N_{f}}%
\bar{\psi}\left( i\gamma ^{\mu }\partial _{\mu }-g\gamma ^{\mu
}A_{\mu }^{a}%
\frac{\lambda ^{a}}{2}-m_{f}\right) \psi,  \label{s1}
\eea
where the subscript $f$ denotes the various quark flavors $u$,
$d$, $s$, $c$ etc., $g$ is the  coupling constant, and $A_{\mu
}^a$ is the vector potential taking values in the Lie algebra with
generators $\lambda ^a$. The nonlinear gluon field strength
$F_{\mu \nu }^{a}$ is given by
\begin{equation}
F_{\mu \nu }^{a}=\partial _{\mu }A_{\nu }^{a}-\partial _{\nu
}A_{\mu
}^{a}+gf_{abc}A_{\mu }^{b}A_{\nu }^{c}\,.  \label{s2}
\end{equation}

QCD predicts a weakening of the quark-quark interaction at short
distances (or high momenta $Q^{2}$), because the one-loop series
for the gluon propagator yields a running coupling constant
\cite{3}
\begin{equation}
g^{2}\left( Q^{2}\right) =\frac{16\pi ^{2}}{\left(
11-2N_{f}/3\right) \ln
\left| Q^{2}/\Lambda ^{2}\right| },  \label{s3}
\end{equation}
where $N_{f\text{ }}$ is the number of active quark
flavors and the QCD scale parameter $\Lambda \approx 200$ MeV. The
coupling constant $g^{2}\left( Q^{2}\right) $ vanishes for high
momenta $Q^{2}$, and tends to infinity for $N_f\rightarrow 33/2$.

\subsection{The MIT bag model equation of state}

Assuming that interactions of quarks and gluons are sufficiently
small, the energy density $\varepsilon $ and pressure $P$ of a
quark-gluon plasma at temperature $T$ and chemical potential $\mu
_{f\text{ }}$can be calculated by thermal theory. Neglecting quark
masses in first order perturbation theory, the equation of state
is \cite{3}
\begin{eqnarray}
\varepsilon &=&\left( 1-\frac{15}{4\pi }\alpha _{s}\right)
\frac{8\pi ^{2}}{15}T^{4}+N_{f}\left( 1-\frac{50}{21\pi
}\alpha_{s}\right) \frac{7\pi ^{2}}{10} T^{4}
    \nonumber\\
&&+ \sum_{f}3\left( 1-2\frac{\alpha _{s}}{\pi }\right) \left( \pi
^{2}T^{2}+\frac{\mu
_{f}^{2}}{2}\right) \frac{\mu _{f}^{2}}{\pi ^{2}}+B,  \label{s4}
\end{eqnarray}
or
\begin{equation}
\varepsilon =\sum_{i=u,d,s,c;e^{-},\mu ^{-}}\varepsilon _{i}+B,
\label{s5}
\end{equation}%
where $\alpha _s$ is the strong interaction coupling constant, and
$B$ is the difference between the energy density of the
perturbative and non-perturbative QCD vacuum (the bag constant).
The thermodynamic parameters of the quark-gluon plasma are related by the equation of state of the quark matter,  given by
\begin{equation}
P=\frac{1}{3}\left( \varepsilon -4B\right),  \label{s6}
\end{equation}
or
\begin{equation}
P+B=\sum_{i=u,d,s,c;e^{-},\mu ^{-}}p_{i}.  \label{s7}
\end{equation}

The entropy density of the quark-gluon plasma is given by
$s=\left( \partial p/\partial T\right) _{\mu }$. Equation
(6) is essentially the equation of state of a gas of massless
particles with corrections due to the QCD trace anomaly and
perturbative interactions. These are always negative, and when
$\alpha _{s}=0.5$ they reduce the
energy density at a given temperature by about a factor of two \cite{3}.

Most of the investigations of the stellar quark-gluon plasma have
been done under the assumption of the electric charge neutrality
of the quark-gluon plasma that reads $\sum_{i=u,d,s;e^{-},\mu
^{-}}q_{i}n_{i}=0$. In the case of a star formed from massless
$u$, $d$ and $s$ quarks the charge neutrality condition can be
explicitly formulated as $2n_{u}/3=\left(n_{d}+n_{s}\right) /3$
\cite{2}.

More sophisticated investigations of quark-gluon interactions have
shown that Eq.~(\ref{s6}) represents a limiting case of more
general equations of state. For example, MIT bag models with
massive strange quarks and lowest order QCD interactions lead to
some correction terms in the equation of state of quark matter.
Models incorporating restoration of chiral quark masses at high
densities and giving absolutely stable strange matter can no
longer be accurately described by using Eq.~(\ref{s6}). If the
quark interaction is described by a colour-Debye-screened
inter-quark vector potential, originating from gluon exchange, and
by a density-dependent scalar potential, which restores chiral
symmetry at high density (in the limit of massless quarks) the
resulting EOS has asymptotic freedom built in, shows confinement
at zero baryon density, and deconfinement at high density. This
density-dependent scalar potential arises from the density
dependence of the in-medium effective quark masses $m_q$, which
are assumed to depend on the baryon number density $n_B$
\cite{De98}.

On the other hand, in these types of models the equation of state
$P=P\left( \varepsilon \right) $ can be well approximated by a
linear function in the energy density $\varepsilon $ \cite{Go00}.
The linear approximation of the equation of state was studied in
\cite{Zd00}, and all the parameters of the EOS have been obtained
as polynomial functions of the strange quark mass, QCD coupling
constant and bag constant.

\subsection{Color Flavor Locked quark matter}

It is generally agreed today that the Color-Flavor-Locked  state
is likely to be the ground state of matter, at least for
asymptotic densities, and even if the quark masses are unequal
\cite{Al1,Al2,Al3,Al4,cfl2}. Moreover, the equal number of flavors
is enforced by symmetry, and electrons are absent, since the
mixture is automatically neutral. The properties of the  CFL quark
matter  depends strongly on the values of the deconfinement phase
transition density and the CFL gap parameter, which are poorly
known from both a theoretical and experimental point of view. The
free energy density $\Omega _{CFL}$ for quark matter in the CFL
phase is given by \cite{42}
\bea
\Omega _{CFL}\left(\mu ,
\mu_e\right)&=&\Omega_{CFL}^{quarks}\left(\mu \right)+\Omega
_{CFL}^{GB}\left(\mu ,\mu _e \right)\nonumber\\
&&+\Omega _{CFL}^{electrons}\left(\mu _e\right),
\eea
where $\Omega _{CFL}^{GB}$ is the contribution from the Goldstone
bosons arising due to the breaking of chiral symmetry in the CFL
phase. By assuming that the mass $m_{s}$ of the $s$ quark is not
large compared to the chemical potential $\mu $, the
thermodynamical potential of the quark matter in the CFL phase can
be approximated as \cite{LuHo02}
\bea
\Omega _{CFL}&=&-\frac{3\mu ^{4}}{4\pi ^{2}}+\frac{3m_{s}^{2}}
{4\pi ^{2}}-%
\frac{1-12\ln \left( m_{s}/2\mu \right) }{32\pi ^{2}}m_{s}^{4}
    \nonumber\\
&&-\frac{3}{\pi
^{2}}\Delta ^{2}\mu ^{2}+B,
\eea
where $\Delta $ is the gap energy. With the use of this expression
the pressure $P$ of the quark matter in the CFL phase can be
obtained as an explicit function of the energy density
$\varepsilon $ in the form \citep%
{LuHo02}
\begin{equation}\label{pres}
P=\frac{1}{3}\left( \varepsilon -4B\right) +\frac{2\Delta
^{2}\delta ^{2}}{\pi
^{2}}-\frac{m_{s}^{2}\delta ^{2}}{2\pi ^{2}},
\end{equation}
where
\begin{equation}
\delta ^{2}=-\alpha +\sqrt{\alpha ^{2}+\frac{4}{9}\pi ^{2}\left(
\varepsilon
-B\right) },
\end{equation}%
and
\be\label{alpha}
\alpha =-\frac{m_{s}^{2}}{6}+\frac{2\Delta ^{2}}{3}.
\ee

Thus, Eq. (\ref{pres}) can finally be expressed as
\begin{equation}\label{pres2}
P=\frac{1}{3}\left( \varepsilon -4B\right) +\frac{3 \alpha \delta
^{2}}{\pi^{2}},
\end{equation}
which will be useful in the analysis outlined below.

\section{Field equations for static and spherically symmetric wormholes}\label{secII}

In this work, motivated by the proposal that at high densities the phases of quark matter are
described either by the equation of state  of the  MIT bag model, or
by the CFL phase equation of state, we
consider the possibility that wormhole geometries can be supported
by quark matter, in both normal and
superconducting states.

In the following we  assume that the  wormhole metric takes
the form \cite{Morris}
\begin{equation}
ds^{2}=-e^{2\Phi(r)}dt^{2}+\frac{dr^{2}}{1-b(r)/r}+r^{2}(d\theta
^{2}+\sin
^{2}\theta d\phi ^{2})\,,
  \label{WHmetric}
\end{equation}
where the metric function $\Phi(r)$ is denoted the redshift
function and $b(r)$ the shape function \cite{Morris}.
The redshift function $\Phi(r)$ must be finite everywhere to avoid
the presence of event horizons \cite{Morris}. In order to have a
wormhole geometry, the shape function $b(r)$ must obey the flaring
out condition of the throat, which translates as $(b-
b^{\prime}r)/b^{2}>0$ \cite{Morris}. At the throat, we have
$b(r_{0})=r=r_{0}$, and taking into account the flaring-out condition the inequality $b^{\prime}(r_{0})<1$ is
imposed.

In classical general relativity, taking into account the
above-mentioned flaring-out condition, and through the Einstein
field equation one deduces that the matter threading the wormhole
throat violates the null energy condition (NEC). More
specifically, the NEC imposes that $T_{\mu\nu}k^\mu k^\nu \geq 0$,
where $k^\mu$ is {\it any} null vector. Thus, a fundamental
ingredient in wormhole physics, in classical general relativity,
is the violation of the NEC, i.e., $T_{\mu\nu}k^\mu k^\nu < 0$ somewhere more specifically, at the wormhole throat and its vicinity). Matter satisfying the latter condition is denoted as {\it exotic
matter}.

The field equations are given by the following stress-energy
scenario
\begin{eqnarray}
\varepsilon(r)&=&\frac{1}{8\pi} \;\frac{b'}{r^2}
\label{rhoWH1},\\
p_r(r)&=&\frac{1}{8\pi} \left[2 \left(1-\frac{b}{r}
\right) \frac{\Phi'}{r} -\frac{b}{r^3}\right]  \label{prWH1},\\
p_t(r)&=&\frac{1}{8\pi} \left(1-\frac{b}{r}\right)\Big[\Phi ''+
(\Phi')^2- \frac{b'r-b}{2r(r-b)}\Phi' -
   \nonumber  \\
&&\frac{b'r-b}{2r^2(r-b)}+\frac{\Phi'}{r} \Big] \label{ptWH1},
\end{eqnarray}
where the prime denotes the derivative with respect to the radial coordinate $r$,  $\varepsilon(r)$ is the energy density, $p_r(r)$ is the
radial pressure, and $p_t(r)$ is the tangential pressure, measured in
the orthogonal direction to the radial direction, respectively.

Using the conservation of the stress-energy tensor,
$T^{\mu\nu}{}_{;\nu}=0$, we obtain the
following equation
\begin{equation}
p_r'=\frac{2}{r}\,(p_t-p_r)-(\varepsilon +p_r)\,\Phi'
\label{prderivative} \,,
\end{equation}
which can be interpreted as the relativistic Euler equation, or
the hydrostatic equation for equilibrium for the material
threading the wormhole.

Note that now we have three independent equations, Eqs.~(\ref{rhoWH1})-(\ref{ptWH1}), with five unknown functions of the
radial coordinate $r$, i.e., $\Phi(r)$, $b(r)$, $\varepsilon(r)$,
$p_r(r)$ and $p_t(r)$. To solve the system, different strategies
have been adopted in the literature. For instance, one may
model an appropriate spacetime geometry by considering a specific
equation of state and impose one of the functions $\Phi(r)$ or
$b(r)$, thus closing the system of the coupled differential
equations. One may also impose the form of the functions $b(r)$ and $\Phi(r)$ by hand and
consequently determine the stress-energy tensor components. Conversely, one could construct a suitable source for the spacetime
geometry by imposing the stress-energy components, and
consequently determine the metric fields.

In this work, we consider a variant of the first approach, by
choosing one of the quark model equations of state, and exploring
specific functions of the radial coordinate that appear in the
resulting differential equations, to find specific wormhole solutions.

\section{Specific solutions: Wormhole geometries supported by the MIT bag model
equation of state}

Despite the fact that the MIT bag model equation of state
represents an isotropic pressure, in the context of quark compact
spheres, instability inhomogeneities may form as a result of density perturbations. Therefore, the pressure in the
MIT bag model equation of state may be regarded a radial pressure,
and the tangential pressure is determined through the Einstein
field equations.

Thus, taking into account the MIT bag model equation of state,
given in the form
\begin{equation}
p_r=\frac{1}{3}\left( \varepsilon -4B\right),  \label{MITbag2}
\end{equation}
and using the Eqs.~(\ref{rhoWH1})--(\ref{prWH1}), we deduce the
following differential equation
\begin{equation}
\Phi ^{\prime }(r)=\frac{r}{2\left[ 1-b(r)/r\right] }\left[
\frac{b(r)}{r^{3}%
}+\frac{b^{\prime }(r)}{3r^{2}}-\frac{32\pi }{3}B(r)\right]
\label{ode1}\,.
\end{equation}
Due to the high energy density regime considered in the Introduction, we assume that the factor $B$, which is the difference
between the energy density of the perturbative and
non-perturbative QCD vacuum, is a function of the radial
coordinate, i.e., $B=B(r)$.

\subsection{Constant MIT bag parameter}

In this section, we consider a constant MIT bag parameter, i.e., $B=B_0$, in order to gain some insight into the physics involved.

\subsubsection{Constant redshift function, $\Phi'(r)=0$}

First, consider a constant redshift function, $\Phi'(r)=0$, so that the differential equation, Eq. (\ref{ode1}) yields the following solution for the shape function
\begin{equation}
b(r)=\frac{16}{3}\pi B r^3 \left[ 1 - \left( \frac{r_0}{r} \right)^6 \right] + r_0 \left( \frac{r_0}{r} \right)^3 \,.
   \label{defshapecase1}
\end{equation}

Note that this solution is not asymptotically flat, so that it needs to be matched to an exterior vacuum solution.
For instance, consider that the exterior solution is the
Schwarzschild spacetime, given by
\begin{eqnarray}
ds^2&=&-\left(1-\frac{2M}{r}\right)\,dt^2+
\left(1-\frac{2M}{r}\right)^{-1}dr^2
   \nonumber  \\
&&+r^2(d\theta ^2+\sin
^2{\theta}\, d\phi ^2) \label{metricvacuumlambda}  \,.
\end{eqnarray}
In this case the spacetimes given by the metrics Eqs. (\ref{WHmetric}) and (\ref{metricvacuumlambda}) are matched at $a$, and one has a thin shell surrounding the wormhole. Using the Darmois-Israel formalism \cite{thinshell}, the surface stresses are given by
\begin{eqnarray}
\sigma&=&-\frac{1}{4\pi a} \left(\sqrt{1-\frac{2M}{a}}-
\sqrt{1-\frac{b(a)}{a}} \, \right)
    \label{surfenergy}   ,\\
{\cal P}&=&\frac{1}{8\pi a}
\left(\frac{1-\frac{M}{a}}{\sqrt{1-\frac{2M}{a}}}- [1+a\Phi'(a)] \,
\sqrt{1-\frac{b(a)}{a}} \, \right)
    \label{surfpressure}    ,
\end{eqnarray}
where $\sigma$ is the surface energy density and ${\cal
P}$ the surface pressure.

The surface mass of the thin shell is given by $m_s=4\pi a^2 \sigma$, namely,
\begin{eqnarray}
m_s = a \left(\sqrt{1-\frac{b(a)}{a}}  -\sqrt{1-\frac{2M}{a}} \, \right)
    \label{surfmass}     .
\end{eqnarray}
If one imposes a positive surface mass of the thin shell, $m_s>0$, then the condition $b(a)<2M$ follows.

Furthermore, one may interpret $M$ as the total mass of the system, given by
\begin{eqnarray}
M = \frac{b(a)}{2} +m_s \left(\sqrt{1-\frac{b(a)}{a}}  -\frac{m_s}{2a} \, \right)
    \label{totalmass}     \,,
\end{eqnarray}
which in this case is the total mass of the wormhole in one asymptotic region.

Taking into account the flaring-out condition at the throat, $b'(r_0)<1$, one arrives at the restriction $8\pi B r_0^2<1$, which places an upper bound on the wormhole throat
\begin{equation}
r_0^2 < \frac{1}{8\pi B}\,.
\end{equation}
Using the plausible values for the bag constant, such as $B=56 \, {\rm Mev/fm}^3$, provided in \cite{2c}, we immediately find an upper bound on the throat radius given by $r_0 \lesssim 10^4 \,{\rm m}$. Thus, this bound is in excellent agreement with the macroscopic wormholes theoretically constructed in this work.

The stress-energy profile for this case, taking into account $\Phi'(r)=0$ and Eq. (\ref{defshapecase1}), is given by the following expressions
\begin{eqnarray}
\varepsilon(r)&=&\frac{1}{8\pi} \;\frac{16 \pi B (r^6+r_0^6)-3r_0^4}{r^6}
\label{rhoWH1case},\\
p_r(r)&=&\frac{1}{24\pi} \frac{16 \pi B (r_0^6-r^6)-3r_0^4}{r^6}  \label{prWH1case},\\
p_t(r)&=&\frac{1}{12\pi} \frac{8 \pi B (r^6+2r_0^6)-3r_0^4}{r^6}\label{ptWH1case},
\end{eqnarray}
which is finite throughout the interior range $r_0\leq r \leq a$.

\subsubsection{Isotropic pressure: $p_r(r)=p_t(r)$}

Consider the case of isotropic pressure, $p_r(r)=p_t(r)=p(r)$, so that the conservation equation reduces to $p'(r)=-[\epsilon(r) + p(r) ]\,\Phi'(r)$, and yields the solution
\begin{equation}
p(r)=-B+Ce^{ -4\Phi(r) }\,.
\end{equation}
The integration constant $C$ is given by $C=(p_0+B)e^{4\Phi_0}$, where $p_0$ and $\Phi_0$ are the values of the pressure and redshift function evaluated at the throat. Thus, the isotropic pressure is finally given by
\begin{equation}
p(r)=-B+(p_0+B)e^{ -4[\Phi(r)-\Phi_0 ]}\,.
\end{equation}

From the field equation (\ref{rhoWH1}), one deduces the relationship
\begin{eqnarray}
\Phi(r)&=&\Phi_0-\frac{1}{4} \ln \left\{  \frac{1}{3(p_0+B)} \left[ \frac{b'(r)}{8\pi r^2} - B \right] \right\}
   \nonumber \\
&=&-\frac{1}{4} \ln \left\{  \frac{e^{-4\Phi_0}}{3(p_0+B)} \left[ \frac{b'(r)}{8\pi r^2} - B \right] \right\}\,.
  \label{isoPhi}
\end{eqnarray}

From Eq. (\ref{isoPhi}), one obtains the following generic restriction $b'(r)/(8\pi r^2)-B > 0$, which at the throat reduces to  $b'_0 > 8\pi r_0^2 B$. From the flaring-out condition at the throat, $b'(r_0)<1$, one obtains the upper bound on the wormhole throat $r_0^2 < 1/(8\pi  B)$.

\subsubsection{Specific radial coordinate-dependent bag function}

A particular solution may also be deduced by considering a
constant redshift function, i.e., $\Phi'=0$, and specifying the
following choice for the bag function
\begin{equation}
\frac{32\pi}{3}B(r)=\frac{1}{ r_0^2}\left(\frac{r_0}{r}\right)^n \,.
\end{equation}
By taking into account the differential equation (\ref{ode1}),
one finally ends up with the following shape function
\begin{equation}
b(r)=\frac{r_0}{6-n}\left[3\left(\frac{r_0}
{r}\right)^{n-3}+\left(3-n\right)\left(\frac{r_0}{r}\right)^3 \right] \,,
\end{equation}
which is always positive and asymptotically flat for $2\leq n\leq 3$. This solution satisfies the condition $b'\left(r_0\right)=0$. The flaring out condition $\left[b(r)-rb'(r)\right]/b^2>0$ gives the condition
\begin{equation}
\frac{4(n-3)-3(n-2)\left(r/r_0\right)^{n-6}}{n-6}>0.
\end{equation}

The stress-energy tensor components for this solution are given by
\begin{equation}
\rho(r)=\frac{3}{8\pi r_0^2}\left( \frac{n-3}{n-6} \right) \, \left[ \left( \frac{r_0}{r} \right)^n -\left( \frac{r_0}{r} \right)^6  \right],
\end{equation}
\begin{equation}
p_r(r)=-\frac{1}{8\pi r_0^2} \left( \frac{1}{6-n} \right) \, \left[ 3\left( \frac{r_0}{r} \right)^n -(3-n)\left( \frac{r_0}{r} \right)^6  \right],
\end{equation}
\begin{eqnarray}
p_t(r)=\frac{1}{16\pi r_0^2} \left( \frac{1}{6-n} \right)
   \Bigg[ 3(2-n)\left( \frac{r_0}{r} \right)^n
   \nonumber \\
   +4(n-3)\left( \frac{r_0}{r} \right)^6  \Bigg]\,,
\end{eqnarray}
which are finite throughout the spacetime geometry. Note that the energy density is zero at the throat, and $p_r=-1/(8\pi r_0^2)$ as expected.

\subsection{Shape function dependent bag function}

A careful analysis of the solutions to the differential equation
(\ref{ode1}), shows several problematic issues related to wormhole
physics. First, considering a specific shape function, $b(r)$, one
immediately verifies that solving Eq.~(\ref{ode1}) for $\Phi(r)$
produces solutions with event horizons, i.e., $\Phi(r) \propto
\ln(1-b(r)/r)$, rendering the wormhole non-traversable. This
difficulty arises due to the factor $(1-b(r)/r)$ in the
denominator in Eq. (\ref{ode1}).

Now, in order to avoid the presence of event horizons, one may choose a suitable bag function $B(r)$ of the form
\begin{equation}
\frac{32\pi}{3}B(r)= \frac{b(r)}{r^{3}}+\frac{b^{\prime }
(r)}{3r^{2}}-\left[ 1-\frac{b(r)}{r}\right]\frac{C_0}
{r_0^{2}}\left(\frac{r_0}{r}\right)^n  .
   \label{Bagfunc}
\end{equation}
By substituting this choice into the differential equation Eq.~(\ref{ode1}),
one finds the following solution for the redshift function
\begin{equation}
\Phi(r)=C_1-\frac{C_0}{2(n-2)}\left(\frac{r_0}{r}\right)^{n-2},
  \label{redshift5}
\end{equation}
where $C_0$ is an arbitrary constant, and $n \neq 2$, and $C_1$ is an integration constant which can be reabsorbed in a redefinition of the time coordinate, so we set $C_1=0$, without a loss of generality. For $n>2$, the redshift function is finite for all $r$, falls off to zero as $r \rightarrow \infty$. For the specific solution of $n=2$, we have the following logarithmic solution, $\Phi(r)=C_1+\frac{C_0}{2}\ln(r)$, which we exclude from the analysis.

We emphasize that this solution has the feature that one could leave the function $b(r)$ generic. However, one may suitably model a wormhole geometry by specifically choosing the shape function, which consequently also specifies the bag function given by Eq. (\ref{Bagfunc}). More specifically, note that the introduction of a radial-dependent Bag function, introduces a new unknown function so that we are left with a new degree of freedom, so for instance, so that as mentioned above we may choose a specific shape function.

In this context, consider the particular choice of the shape function
$b(r)=r_0(r/r_0)^\alpha$, with $0<\alpha<1$. For this case we
readily verify that $b'(r)=\alpha(r/r_0)^{\alpha-1}$, so that at
the throat $b'(r_0)=\alpha<1$, and that for $r\rightarrow \infty$
we have $b(r)/r=(r_0/r)^{1-\alpha}\;\rightarrow 0$.
In addition to this choice for the shape function, consider the redshift function given above by Eq. (\ref{redshift5}), but re-written as $\Phi(r)=\Phi_0(r_0/r)^\beta$, with $\beta=n-2 > 0$ and $\Phi_0=-C_0/2(n-2)$. Note that this choice of the redshift function is finite everywhere, so that no event horizons are present.

Thus, Eq. (\ref{MITbag2}) provides the following Bag function
\begin{eqnarray}
B(r)&=& \frac{3}{32\pi r^2} \Bigg \{\left(1+\frac{\alpha}{3} \right)\left(\frac{r_0}{r}\right)^{1-\alpha}
    \nonumber \\
&&+ 2\beta \Phi_0 \left(\frac{r_0}{r}\right)^{\beta}\left[1-\left(\frac{r_0}{r}\right)^{1-\alpha} \right]  \Bigg \}\,,
   \label{sol1Bagfunc}
\end{eqnarray}
which reduces to $B(r_0)=B_0=3(1+\alpha/3)/(32\pi r_0^2)$ at the wormhole throat and $B \rightarrow 0$ for $r\rightarrow \infty$.

The stress-energy profile is given by
\begin{eqnarray}
\varepsilon(r)&=&\frac{1}{8\pi r_0^2} \;\left(\frac{r_0}{r}\right)^{3-\alpha}
\label{rhoWH2case},\\
p_r(r)&=&\frac{1}{\pi r_0^2} \; \Bigg\{ 2\beta \Phi_0 \left[1-\left(\frac{r_0}{r}\right)^{1-\alpha} \right]
\left(\frac{r_0}{r}\right)^{2+\beta}
   \nonumber  \\
&&-\left(\frac{r_0}{r}\right)^{3-\alpha}   \Bigg\}
\label{prWH2case},\\
p_t(r)&=&\frac{1}{12\pi} \frac{8 \pi B (r^6+2r_0^6)-3r_0^4}{r^6}\label{ptWH2case},
\end{eqnarray}
which is finite throughout the interior range $r_0\leq r \leq a$.


It is also interesting to consider the ``volume integral
quantifier,'' which provides information on the total amount of
matter violating the averaged null energy condition in the
spacetime. This is defined by $I_V=\int[\varepsilon(r)+p_r(r)]dV$ (see
Ref. \cite{Visser:2003yf} for details), and with a cut-off of the
stress-energy at $a$ is given by
\begin{equation}
I_V=\int_{r_0}^a
(r-b)\left[\ln\left(\frac{e^{2\Phi}}{1-b/r}\right)\right]'\,dr \,.
            \label{Iv}
\end{equation}
Taking into account the shape and redshift functions provided above, the ``volume integral
quantifier'' given by Eq. (\ref{Iv}) provides the following solution
\begin{eqnarray}
I_V&=& \frac{r_0}{\alpha(\beta-\alpha)(1-\beta)}\times
    \nonumber \\
&&  \hspace{-1.15cm}
\Bigg\{ \left[ \left( \frac{a}{r_0} \right)^{\alpha} -1 \right] (\alpha-\beta) \times
\left[ 1+ \alpha \beta -(\alpha + \beta)   \right] +2\alpha \beta \Phi_0 \times
   \nonumber \\
&& \hspace{-0.6cm}  \Bigg[1-\alpha +(\alpha - \beta)\left( \frac{a}{r_0} \right)^{\alpha}
    +(\beta -1)\left( \frac{a}{r_0} \right)^{\alpha - \beta} \Bigg] \Bigg\} \,.
\end{eqnarray}
Now taking the limit $a\rightarrow r_0$, one verifies that $I_V\rightarrow 0$. Therefore, as in the examples presented in Refs. \cite{phantomWH,Visser:2003yf}, one verifies that, in principle, one may construct wormhole geometries with vanishingly small amounts of quark matter violating the averaged null energy condition.\\


An interesting constraint on the wormhole's dimensions, in
particular, on the throat radius may be inferred from the tidal
acceleration restrictions \cite{Morris}. The latter
constraints as measured by a traveler moving radially through the
wormhole, are given by the following inequalities
\begin{eqnarray}
\left |\left (1-\frac{b}{r} \right ) \left [\Phi ''+(\Phi ')^2-
\frac{b'r-b}{2r(r-b)}\Phi' \right] \right
|\,\big|\eta^{\hat{1}'}\big| \leq  g_\oplus   \,,
    \label{radialtidalconstraint}
\end{eqnarray}
\begin{eqnarray}
\left | \frac{\gamma ^2}{2r^2} \left [v^2\left (b'-\frac{b}{r}
\right )+2(r-b)\Phi ' \right] \right | \,\big|\eta^{\hat{2}'}\big|
\leq   g_\oplus    \,, \label{lateraltidalconstraint}
\end{eqnarray}
where $g_\oplus$ is the Earth's gravitational acceleration, and
$\eta^{\hat{i}'}$ is the separation between two arbitrary
parts of his body measured in the traveler's reference frame. We
shall consider $|\eta^{\hat{i}'}|=|\eta|$, for simplicity. We
refer the reader to Ref. \cite{Morris} for details. The
radial tidal constraint, inequality (\ref{radialtidalconstraint}),
constrains the redshift function; and the lateral tidal
constraint, inequality (\ref{lateraltidalconstraint}), constrains
the velocity with which observers traverse the wormhole. These
inequalities are particularly simple at the throat, $r_0$,
\begin{eqnarray}
|\Phi '(r_0)| \leq  \frac{2g_{\oplus}\,r_0}{(1-b')\,|\eta|} \,,
  \qquad
 \gamma^2 v^2 \leq  \frac{2g_{\oplus}\,r_0^2}{(1-b')\,|\eta|}
\label{lateraltidalconstraint2}  \,.
\end{eqnarray}

One may also consider that there exist two space stations positioned
outside the junction radius, $a$, at $l=-l_1$ and $l=l_2$,
respectively, where $dl=(1-b/r)^{-1/2}\,dr$ is the proper radial
distance. Now, the traversal time as measured by an observer traversing through the wormhole and for
the observers that remain at rest at space stations are given by
\begin{eqnarray}
\Delta \tau =\int_{-l_1}^{+l_2} \frac{dl}{v\gamma}
\qquad {\rm and} \qquad
\Delta t =\int_{-l_1}^{+l_2} \frac{dl}{v e^{\Phi}}  \,,
\end{eqnarray}
respectively.

Consider now the wormhole geometry constructed in this section. In addition to this, assume a constant non-relativistic, $\gamma \approx 1$, traversal velocity, and considering the equality cases of (\ref{lateraltidalconstraint2}), we obtain the following relationships
\begin{equation}
r_0^2 \approx  \frac{ \beta \Phi_0 (1-\alpha)  \,|\eta |}{2g_\oplus} \,, \qquad
v \approx
r_0\,\sqrt{\frac{2g_\oplus}{(1-\alpha)\,|\eta |}}
\label{lateraltidalconstraint3}
\end{equation}

For simplicity, we assume that $|\eta| \approx 2 {\rm m}$. From the second restriction of (\ref{lateraltidalconstraint3}), taking into account $\alpha=1/2$, and imposing that the wormhole throat is given by $r_0\approx 10^2\,{\rm m}$, then one obtains $v\approx 4\times 10^2\,{\rm m/s}$ for the traversal velocity. If one considers that the junction radius is given by $a\approx 10^4\,{\rm m}$, then from the traversal times $\Delta \tau \approx \Delta t \approx 2a/v$ (assuming for simplicity that $\Phi \ll 1$), one obtains $\Delta \tau \approx \Delta t \approx 50 \,{\rm s}$.

\subsection{Wormhole geometries supported by a MIT bag model
equation of state with electric charge}

The total stress-energy tensor $T_{\mu }^{\nu }$ inside the
wormhole is assumed to be the sum of two parts $M_{\mu}^{\nu}$ for
the quark matter and $E_{\mu}^{\nu}$ for an electromagnetic
contribution, respectively:
$T_{\mu}^{\nu}=M_{\mu}^{\nu}+E_{\mu}^{\nu}$.

The stress-energy tensor for an anisotropic distribution of quark matter is provided by
\begin{equation}
M_{\mu}^{\nu}=(\epsilon+p_t)u_\mu \, u^\nu+p_t\,
\delta_{\mu}^{\nu}+(p_r-p_t)\chi_\mu \chi^\nu \,,
\end{equation}
where $u^{\mu}=\delta_{0}^{\mu}e^{-\Phi}$ is the
four-velocity satisfying the condition $u_{\mu}u^{\mu}=-1$; $\chi^\mu$ is the unit
spacelike vector in the radial direction, i.e., $\chi^\mu=\sqrt{1-b(r)/r}\,\delta^\mu{}_r$; with $p_r$ and $\epsilon$ related by the bag model equation of state  (\ref{MITbag2}).

The electromagnetic contribution is given by
\begin{equation}
E_{\mu}^{\nu}=\frac{1}{4\pi }\left( F_{\mu \alpha }F^{\nu \alpha}-\frac{1}{4}\delta
_{\mu}^{\nu}F_{\alpha \beta }F^{\alpha \beta}\right) ,
\end{equation}
where $F_{\mu \nu}$ is the electromagnetic field tensor defined in terms
of the four-potential $A_{\mu}$ as
\begin{equation}
F_{\mu \nu}=A_{\mu,\nu}-A_{\nu, \mu},
\end{equation}
where a comma denotes the derivative with respect to the coordinates. For the electromagnetic field we shall adopt the gauge $A_{\mu}=\left( \varphi
\left( r\right) ,0,0,0\right) $.

The Maxwell equations describing the interior of a charged quark
wormhole can be expressed as
\begin{equation}
F_{\mu \nu ,\lambda}+F_{\lambda \mu,\nu }+F_{\nu \lambda ,\mu }=0,
\qquad \nabla _{\nu} F^{\mu \nu}=-\frac{j^{\mu }}{2},
\label{field}
\end{equation}
where $j^{\mu }=\bar{\rho}_{e}u^{\mu }$ is the four-current
density and $\bar{\rho}_{e}$ is the proper charge density. The second equation can be rewritten as:
\begin{equation}
\frac{d}{dr}\left( r^{2}E\right) =\frac{1}{2}\rho _{e}r^{2}.
\label{field4}
\end{equation}

In Eq. (\ref{field4}) $E$ is the usual electric field
intensity defined as $E^{2}=-F_{01}F^{01}$ and $E\left( r\right)
=\left[e^{-
\Phi }\sqrt{1-b(r)/r}\right]\varphi ^{\prime}\left( r\right) $,
with $\varphi^{\prime}\left( r\right) $ =$F_{01}$. The charge
density $\rho _{e}$ in Eq. (\ref{field4}) is related to the proper
charge density $\bar{\rho}_{e}$ by
$\rho_{e}=\bar{\rho}_{e}/\sqrt{1-b(r)/r}$. By integrating
Eq.~(\ref{field4}), we obtain
\be
E\left( r\right) =\frac{q\left(r\right)}{r^{2}},
\ee
where
\bea
q\left( r\right) &=&\frac{1}{2}\int_{0}^{r}\rho
_{e}r^{2}dr
\nonumber\\
&=&\frac{1}{2}\int_{0}^{r}\bar{\rho}_{e}r^{2}dr/\sqrt{1-b(r)/r},
\eea
is the charge within radius $r$.

In the presence of an electric field the gravitational field equations are given by
the following relationships
\begin{eqnarray}
\varepsilon(r)&=&\frac{1}{8\pi} \;\frac{b'}{r^2}  -E^2
\label{rhoWH},\\
p_r(r)&=&\frac{1}{8\pi} \left[2 \left(1-\frac{b}{r}
\right) \frac{\Phi'}{r} -\frac{b}{r^3}\right] +E^2 \label{prWH},
    \\
p_t(r)&=&\frac{1}{8\pi} \left(1-\frac{b}{r}\right)\Big[\Phi ''+
(\Phi')^2- \frac{b'r-b}{2r(r-b)}\Phi'
   \nonumber  \\
&&-\frac{b'r-b}{2r^2(r-b)}+\frac{\Phi'}{r} \Big] -E^2
\label{ptWH}.
\end{eqnarray}

Using the MIT bag model with the equation of state
(\ref{MITbag2}), from Eqs.~(\ref{rhoWH}) and (\ref{prWH}) one arrives at the following
 differential equation
\begin{equation}
\Phi ^{\prime }=\frac{r}{2\left[ 1-b/r\right] }\left[
\frac{b}{r^{3}%
}+\frac{b^{\prime }}{3r^{2}}-\frac{32\pi }{3}B-\frac{32\pi }
{3}\frac{%
q^{2}}{r^{4}}\right] .
\end{equation}

As in the previous example, in order to avoid the presence of
event horizons, we consider the bag function $B(r)$ given by
\begin{equation}\label{32}
\frac{32\pi}{3}B(r)= \frac{b(r)}{r^{3}}+\frac{b^{\prime }(r)}{3r^{2}}%
-\frac{r_0}{r^{3}}\left[ 1-\frac{b(r)}{r}\right]  .
\end{equation}
Analogously, the charge distribution  $q^2(r)$ is taken as $q^2 \propto (1-b/r)$. Taking into account the following choice
\begin{equation}
q^2(r)=q_0r_0^2\left[1-\frac{b(r)}
{r}\right]\left(\frac{r_0}{r}\right)^n \,,
\end{equation}
with $n>0$ and with $q_0$ an arbitrary constant, the solution for the redshift function is given by
\begin{equation}
\Phi(r)=-\frac{r_0}{2r}+\frac{16\pi q_0}{3\left(n+2\right)}\left(\frac{r_0}{r}\right)^{n+2} \,,
\end{equation}
which is finite throughout the radial coordinate range, i.e., $r_0 \leq r < \infty$.

The general solution of Eq.~(\ref{32}) can be obtained as
\begin{eqnarray}
b(r)&=&\frac{e^{3r_0/r}}{r^3}\times \nonumber\\
&&\left[\int{\left(32\pi r^3B(r)+3r_0\right)r^2e^{-3r_0/r}dr+C_2}\right],\nonumber\\
\end{eqnarray}
where $C_2$ is an arbitrary constant of integration.

One may also choose a specific shape function and solve the above equation, or equivalently, use Eq. (\ref{32}) to solve for $B(r)$. For instance, let us consider the
following shape function
\begin{equation}
b(r)=r_0\left( \frac{r}{r_0} \right)^\alpha \,,
\end{equation}
with $0 \leq \alpha <1$.
This yields
\begin{eqnarray}
B(r)=\frac{3}{32\pi r_0^2}\Bigg\{ \left( \frac{r_0}{r} \right)^{3-\alpha}  \left( 1+\frac{\alpha}{3r_0^2} \right)
   \nonumber  \\
 -\left( \frac{r_0}{r} \right)^3 \left[ 1- \left( \frac{r_0}{r} \right)^{1-\alpha} \right]  \Bigg\}.
\end{eqnarray}
The other quantities are given by
\begin{eqnarray}
q^2(r)= q_0 r_0^2 \left( \frac{r_0}{r} \right)^n \left[ 1- \left( \frac{r_0}{r} \right)^{1-\alpha} \right] ,
\end{eqnarray}
 which provides the following expression for the electric field
 \begin{eqnarray}
E^2(r)= \frac{q_0}{ r_0^2} \left( \frac{r_0}{r} \right)^{n+4} \left[ 1- \left( \frac{r_0}{r} \right)^{1-\alpha} \right] .
\end{eqnarray}
The stress-energy tensor profile is finally given by
 \begin{eqnarray}
\varepsilon(r)= \frac{\alpha}{8\pi r_0^2} \left( \frac{r_0}{r} \right)^{2-\alpha} -  \frac{q_0}{ r_0^2} \left( \frac{r_0}{r} \right)^{n+4} \left[ 1- \left( \frac{r_0}{r} \right)^{1-\alpha} \right] ,
\end{eqnarray}

\begin{widetext}

\begin{eqnarray}
p_r(r) &=& \frac{1}{8\pi r_0^2} \Bigg\{ \left( \frac{r_0}{r} \right)^{3}
\left[ 1- \left( \frac{r_0}{r} \right)^{1-\alpha} \right]
\left[ 1-\frac{32\pi q_0}{3} \left( \frac{r_0}{r} \right)^{n+1}  \right]
 -   \left( \frac{r_0}{r} \right)^{3-\alpha} \Bigg\}
+ \frac{q_0}{ r_0^2} \left( \frac{r_0}{r} \right)^{n+4} \left[ 1- \left( \frac{r_0}{r} \right)^{1-\alpha} \right] ,
\end{eqnarray}
 \begin{eqnarray}
p_t(r) &=& \frac{1}{8\pi r_0^2} \Bigg\{ \left[ 1- \left( \frac{r_0}{r} \right)^{1-\alpha} \right] \Bigg\{ \frac{1}{4}  \left( \frac{r_0}{r} \right)^{4}\;\left[ 1-\frac{32\pi q_0}{3} \left( \frac{r_0}{r} \right)^{n+1}  \right]
\left[ 1-\frac{32\pi q_0}{3} \left( \frac{r_0}{r} \right)^{n+1}  + \frac{2r}{r_0} \right]
      \nonumber   \\
 &&      +  \left( \frac{r_0}{r} \right)^{3} \left[ \frac{16\pi q_0 (n+3)}{3}  \left( \frac{r_0}{r} \right)^{2} -1 \right]
+\frac{\frac{1-\alpha}{2} \left( \frac{r_0}{r} \right)^{3-\alpha} }{1-\left( \frac{r_0}{r} \right)^{1-\alpha}}
\left[ 1+ \left( \frac{r_0}{2r} \right)
\left[ 1-\frac{32\pi q_0}{3} \left( \frac{r_0}{r} \right)^{n+1}  \right]
   \right]    \Bigg\}\Bigg\}
 \nonumber   \\
 &&
- \frac{q_0}{ r_0^2} \left( \frac{r_0}{r} \right)^{n+4} \left[ 1- \left( \frac{r_0}{r} \right)^{1-\alpha} \right]  ,
\end{eqnarray}
 which are finite throughout the spacetime.
\end{widetext}


\section{Wormhole geometries supported by Color Flavor Locked
superconducting quark matter}


In this section, we consider wormhole geometries supported by Color Flavor Locked
superconducting quark matter. Now, consider the equation of state given by Eq.~(\ref{pres}), with the
use of Eqs.~(\ref{rhoWH1}) and (\ref{prWH1}) we obtain the
following differential equation
\begin{eqnarray}
2\left[ 1-\frac{b(r)}{r} \right] \frac{\Phi'(r)}{r} - \frac{b(r)}{r^3}-\frac{b'(r)}{3r^2}+\frac{32\pi}{3}B(r)
  \nonumber   \\
+\frac{24}{\pi}\alpha \left\{\alpha -\sqrt{\alpha ^2+\frac{4}{9}\pi ^2 \left[\varepsilon -B(r)\right]}\right\}=0. \label{34}
\end{eqnarray}

We note that the differential equation (\ref{34}) can be solved by
separating terms. For instance, consider the following simplifying assumptions
\begin{eqnarray}
2\left[ 1-\frac{b(r)}{r} \right] \frac{\Phi'(r)}{r} - \frac{b(r)}{r^3}-\frac{b'(r)}{3r^2}
  \nonumber   \\
+\frac{32\pi}{3}B(r)+\frac{24}{\pi}\alpha ^2=0
    \,,
     \label{ode1d}
\end{eqnarray}
and the condition
\begin{equation}\label{56i}
-\frac{24}{\pi}\alpha  \times
   \sqrt{\alpha ^2
+ \frac{4}{9}\pi ^2\left[\varepsilon -B(r) \right]} =0,
 \label{ode1e}
\end{equation}
respectively.

Now, Eq. ( \ref{ode1d}) can be rewritten as
\begin{eqnarray}
\Phi'(r)=
 \frac{r}{2}\left[ 1-\frac{b(r)}{r} \right]^{-1} \Bigg[
 \frac{b(r)}{r^3}+\frac{b'(r)}{3r^2}
  \nonumber   \\
-\frac{32\pi}{3}B(r)-\frac{24}{\pi}\alpha ^2 \Bigg].
  \label{ode1f}
\end{eqnarray}
As before, we verify that a careful analysis of the solutions to Eq. (\ref{ode1f}) shows that considering a specific shape function, $b(r)$, one
immediately verifies that solving the differential equation for $\Phi(r)$
produces solutions with event horizons, i.e., $\Phi(r) \propto
\ln(1-b(r)/r)$, rendering the wormhole non-traversable. As mentioned above, this
difficulty arises due to the factor $(1-b(r)/r)$ in the
denominator in Eq. (\ref{ode1f}). Thus, to avoid this difficult, we consider the following bag function
\begin{eqnarray}
\frac{32\pi}{3}B(r)&=&-\frac{24}{\pi}\alpha ^2 + \frac{b(r)}
{r^{3}}+\frac{b^{\prime }(r)}{3r^{2}}%
   \nonumber  \\
&&-\left[ 1-\frac{b(r)}{r}\right]\frac{C_1}
{r_0^{2}}\left(\frac{r_0}{r}\right)^n ,\label{Bag3}
\end{eqnarray}
where $C_1$ is an arbitrary constant.  The solution for $\Phi $ is
given by
\begin{equation}
\Phi(r)=-\frac{C_1}{2(n-2)}\left(\frac{r_0}{r}\right)^{n-2} ,
\end{equation}
which is finite for $n>2$.

Now, let us consider the second simplifying assumption, given by Eq. (\ref{ode1e})
Assuming that $\alpha \neq 0$, then the term within the square
root is zero, i.e., we obtain
\begin{equation}\label{40}
\alpha ^2
+ \frac{4}{9}\pi ^2\left[\varepsilon -B(r) \right] =0.
\end{equation}
Now, substituting the bag function given by Eq.~(\ref{Bag3}) into
Eq.~(\ref{40}), yields the following differential equation
\begin{equation}
\frac{b'(r)}{r^2}-\frac{b(r)}{r^3}
+\left[ 1-\frac{b(r)}{r}\right]\frac{C_1}{r_0^{2}}\left(\frac{r_0}
{r}\right)^n
 +\frac{48}{\pi}\alpha ^2=0. \label{ode4}
\end{equation}
Note that one now has a certain freedom in choosing a suitable $\Delta$ function. We consider the following choice
\be
\frac{48}{\pi }\alpha ^2=
\frac{b(r)}{r^3}
-\left[ 1-\frac{b(r)}{r}\right]\frac{C_1}{r_0^{2}}\left(\frac{r_0}
{r}\right)^n +\frac{k}{r_0^2}\left(\frac{r_0 }{r} \right)^{k+3} \,,
\ee
which provides the solution for the shape function given by
\begin{eqnarray}
b(r)=r_0\left(\frac{r_0}{r} \right)^k \,,
  \label{defshape3}
\end{eqnarray}
with $k>0$.

For simplicity, consider $k=0$, and after inserting the functional form of $\alpha $ given by
Eq.~(\ref{alpha}), can be rearranged to give
\begin{eqnarray}\label{44}
\Delta(r)=\frac{1}{2}\sqrt{m_s^2\pm
\sqrt{ \frac{3\pi}{4}
\left[\frac{b(r)}{r^3}
-\left[ 1-\frac{b(r)}{r}\right]\frac{C_1}{r_0^{2}}\left(\frac{r_0}
{r}\right)^n \right] }}.\nonumber\\
\end{eqnarray}
Inserting this choice into the  differential equation Eq.~(\ref{ode4}), one immediately obtains the solution for a constant shape function, $b(r)=r_0$, which can also be immediately verified from Eq. (\ref{defshape3}).

Thus, one may write out the stress-energy tensor profile, for the specific solution, i.e., $b(r)=r_0$ and $\Phi(r)=\Phi_0 (r_0/r)^\beta$, with $\Phi_0=-C_1/(n-2)$ and $\beta=n-2>0$, which is given by the following expressions
\begin{equation}
\rho(r)=0,
\end{equation}
\begin{equation}
p_r(r)=-\frac{1}{8\pi r_0^2}\left[ \left( \frac{r_0}{r} \right)^3 +2\beta \Phi_0 \left( \frac{r_0}{r} \right)^{\beta+2} \left( 1- \frac{r_0}{r}  \right)  \right],
\end{equation}
\begin{eqnarray}
p_t(r)&=&\frac{1}{8\pi r_0^2}\,\left( 1- \frac{r_0}{r}  \right) \Bigg\{\beta (\beta+1) \Phi_0 \left( \frac{r_0}{r} \right)^{\beta+2}
   \nonumber   \\
&& +  \beta^2 \Phi_0^2 \left( \frac{r_0}{r} \right)^{2(\beta+1)} - \beta \Phi_0 \left( \frac{r_0}{r} \right)^{\beta+2}
    \nonumber  \\
&& +   \frac{1}{2(1-r_0/r)}\left( \frac{r_0}{r} \right)^{3} \left[ 1- \Phi_0 \beta \left( \frac{r_0}{r} \right)^{\beta}   \right]
   \Bigg\},
\end{eqnarray}
which is finite throughout the spacetime geometry.

\section{Discussions and final remarks}\label{concl}

The quark structure of baryonic matter is the central paradigm of the present-day elementary particle physics. At very high densities, which can be achieved in the interior of neutron stars, a deconfinement transition can break the baryons into their constitutive components, the quarks, thus leading to the formation of the quark-gluon plasma. Moreover, the strange quark matter,  consisting of a mixture of $u$, $d$ and $s$ quarks, may be  the most energetically favorable state of matter. At high densities quark matter may also undergo a phase transition to a color superconducting state. The thermodynamic properties of the quark matter are well-known from a theoretical point of view, and several equations of state of the dense quark-gluon plasma have been proposed  in the framework of a Quantum Chromodynamical approach,  such as the MIT bag model equation of state and the equations of state of the superconducting Color-Flavor-Locked  phase.

Motivated by these theoretical models, in the present paper we have explored the conditions under which wormhole geometries may be supported by the equations of state considered in the theoretical investigations of quark-gluon interactions. Since quark-gluon plasma can exist only at very high densities, the existence of the quark-gluon wormholes requires quark matter at extremely high densities. In these systems the basic physical parameters describing the properties of the QCD quark-gluon plasma (bag constant, gap energy, quark masses) become effective, density and interaction dependent quantities. It is this specific property of the strong interactions we have used to generate specific mathematical functional forms of the bag function and of the gap function that could make possible the existence of a wormhole geometry supported by a strongly gravitationally confined normal or superconducting quark-gluon plasma.

In the case of the normal quark-gluon plasma, wormhole solutions can be obtained by assuming either a specific dependence of $B$ on the shape function $b$, or some simple functional representations of $B$. In both cases in the limit of large $r$ the bag function tends to zero, $\lim_{r\rightarrow\infty}B=0$, and in this limit the equation of state of the quark matter becomes the radiation type equation of the normal baryonic matter, $p=\varepsilon /3$. Therefore, once the density of the quark matter increases after a deconfinement transition, a density (radial coordinate) dependent bag function could lead to the violation of the null energy condition, with the subsequent generation of a wormhole supported by the quark-gluon plasma. A high intensity electric field with a shape function dependent charge distribution could also play a significant role in the formation of the wormhole.

In the case of the superconducting quark matter the gravitational field equations can be solved by assuming that both the bag function and the gap function are shape function and $s$ quark mass dependent quantities. However, in the large $r$ limit, in order to reobtain the standard baryonic matter equation of state, the condition of the vanishing of the mass of the $s$ quark is also required, $\lim_{r\rightarrow\infty}m_s=0$.  The assumption of a zero asymptotic $u$, $d$ and $s$ quark mass is also frequently used in the study of quark star models \cite{2}.

\section*{Acknowledgments}

FSNL is supported by a Funda\c{c}\~{a}o para a Ci\^{e}ncia e Tecnologia Investigador FCT Research contract, with reference IF/00859/2012, funded by FCT/MCTES (Portugal). FSNL also acknowledges financial support of the Funda\c{c}\~{a}o para a Ci\^{e}ncia e Tecnologia through the grants CERN/FP/123615/2011 and CERN/FP/123618/2011.

\end{document}